\documentclass[5p,times]{elsarticle}

\usepackage{amssymb}
\usepackage{amsthm}
\usepackage{makecell}
\usepackage{amsmath}
\usepackage{color,xcolor}
\newtheorem{remark}{Remark}
\usepackage[utf8]{inputenc}
\usepackage{booktabs, caption, makecell}

\usepackage{threeparttable}
\usepackage{epigraph}
\usepackage{enumitem}
\usepackage{hyperref}
\usepackage{subfigure}
\usepackage{mathtools}
\usepackage{url}
\usepackage{color}
\usepackage{bbm}
\usepackage{amsmath, amssymb, amsfonts, amsthm}
\usepackage{siunitx}
\usepackage[linesnumbered, lined, boxed]{algorithm2e}

\usepackage{multirow}
\usepackage[figuresright]{rotating}
\usepackage{amsmath}

\usepackage{tablefootnote}

\usepackage[markup=blue,deletedmarkup=sout,
addedmarkup=nf, final]{changes}

\begin{document}
\begin{frontmatter}
\title{Learning regime-dependent governing equations: A symbolic decision tree approach}

\author{Ilias Mitrai\fnref{label2}}
\fntext[label2]{Corresponding author}
\ead{imitrai@che.utexas.edu}

\author{Tongjia Liu}
\author{Gabriel E. Sanoja}

\address{McKetta Department of Chemical Engineering, The University of Texas at Austin, Austin, TX 78712, US}

%% Abstract
\begin{abstract}
    Many chemical engineering systems are governed by mechanisms that switch across operating regimes, making the data-driven discovery of regime-dependent governing equations essential for predictive modeling, optimization, and control. We propose symbolic decision trees for the data-driven discovery of regime-dependent governing equations. The method simultaneously learns interpretable splitting conditions to partition the input domain and local governing equations that describe each regime. To improve tractability, both the splitting conditions and governing equations are parametrized using basis functions, resulting in a mixed-integer optimization learning problem. We use the proposed approach to learn hybrid dynamical models and a constitutive equation for the zero-shear viscosity of polymer melts. Symbolic decision trees identify physically interpretable regimes and local governing equations while improving predictive accuracy relative to approaches that learn a single global model or use existing decision tree models. This framework provides an interpretable and generalizable route for discovering regime-dependent models in chemical engineering systems. 
\end{abstract}
%% Keywords
%% Keywords
\begin{keyword}
Model discovery \sep mixed integer optimization \sep machine learning
\end{keyword}

\end{frontmatter}

\section{Introduction}
Several chemical engineering systems exhibit behavior governed by regime-dependent phenomena. For example, dominant transport mechanisms vary across flow regimes, biochemical networks activate or deactivate based on cell state, and model predictive control (MPC) laws change with active constraints. Developing mathematical models for such systems requires discovering governing equations that are specific to each regime.

Standard model discovery approaches typically learn a single model using basis functions \cite{cozad2014learning, brunton2016discovering, narasingam2018data, abdullah2023data, muthyala2025symantic}, symbolic regression \cite{schmidt2009distilling, cozad2018global, a+17, kim2023learning}, and hybrid modeling strategies \cite{bradley2022perspectives}.  However, discovering regime-dependent governing equations requires simultaneously identifying logical conditions that define distinct regimes and the governing equations that describe behavior within each regime.

Several approaches have been proposed to discover such models; their difference lies primarily in the identification of the regimes and the parameterization of the local nonlinear expressions. An approach is to use deep learning, such as neural networks, to learn models including differential equations \cite{chen2020learning, poli2021neural} and MPC policies \cite{kumar2021industrial, hertneck2018learning, russo2023learning, karg2018deep, cauligi2022prism}. Although neural networks can be accurate, they lack interpretability and do not provide direct insight into distinct regimes or the transitions between them. 

An alternative is to partition the learning task into two steps: first, identify the regimes and then fit a model for each regime. In such approaches, the regimes are identified by clustering the data using unsupervised machine learning algorithms \cite{nakada2005identification, mangan2019model}. Clustering-based approaches have two key limitations. First, the clustering step can provide interpretable splitting conditions only for certain classes of systems, such as piecewise-affine dynamical systems, where the state space is partitioned into polyhedral domains \cite{ferrari2003clustering}. Second, regime and model discovery are performed sequentially; thus, the partition is not optimized for model accuracy.

The third approach to regime-dependent governing equation discovery simultaneously identifies the partition and the local models. This approach has been used extensively to learn dynamical models for piecewise-affine dynamical systems, in which the domain partition is polyhedral and the model is linear within each regime \cite{paoletti2007identification, bemporad2001identification, yang2016mathematical}. Conceptually, these methods resemble fitting a decision tree with linear splitting logic and linear expressions at the leaves.

Decision trees with constant \cite{breiman2017classification}, piecewise-linear \cite{quinlan1992learning}, and polynomial \cite{bertsimas2021near} expressions at the leaves have been used recently to approximate MPC laws \cite{drgovna2018approximate, klauvco2014building,masti2020learning, bemporad2022piecewise} and develop models for production planning \cite{wu2024piecewise}. However, their application to nonlinear systems remains challenging, as the governing equations in each regime can be highly nonlinear, and approximating them with hyperplanes often requires a very fine partition to achieve high accuracy, leading to decision trees with many nodes \cite{sunshine2025hyperplane}. This challenge can be mitigated by using nonlinear expressions for both the splitting logic and the local governing equations. 

An approach is to leverage symbolic regression and represent the splitting logic and local governing equations as expression trees \cite{ly2012learning, zhang2022ps, plambeck2025identification, fong2024symbolic}. Although this approach is generic and does not require any a priori assumptions about the form of the branching logic or local nonlinear expressions, the search is computationally expensive. Existing approaches rely on genetic programming \cite{zhang2022ps, plambeck2025identification, fong2024symbolic}, which is a heuristic and cannot guarantee optimality of the learned model with respect to a desired objective function. 

In recent work, we proposed a mixed-integer linear optimization approach for learning piecewise nonlinear MPC laws \cite{mitrai2025discovering}. In this approach, the control law is modeled as a decision tree with axis-aligned splits, meaning that each splitting logic depends on a single variable, while nonlinear expressions at the leaves are parameterized using basis functions. Application of this method to a prototypical isothermal continuously stirred tank reactor showed that it can lead to interpretable and accurate MPC laws with closed-loop performance similar to that of standard MPC. 

In this paper, we propose symbolic decision trees as a general framework for discovering regime-dependent governing equations in chemical engineering systems. The proposed approach simultaneously learns the partition of the input domain into distinct linear and/or nonlinear regimes, the splitting logic that defines transitions between regimes, and the local governing equations within each regime. The splitting logic is modeled with binary variables, following the optimal classification tree formulations presented in \cite{bertsimas2017optimal,aghaei2025strong}. We parameterize both the splitting logic and the local expressions using basis functions, such that the resulting learning task is a mixed integer optimization problem that can be solved with current state-of-the-art optimization algorithms. 

We use the proposed approach to discover regime-dependent models for hybrid dynamical systems and material properties, such as the zero-shear viscosity of polymers. The results show that the proposed approach can recover both the splitting logic and the local governing equations. The rest of the paper is organized as follows: In Section~\ref{sec: model} we present the learning model, in Section~\ref{sec: simple example} we apply the proposed approach to a prototypical case study, in Section~\ref{sec: learn hybrid models} we consider the learning of hybrid models, in Section~\ref{sec: predict polymer property} we learn regime-dependent models for polymer properties.

\section{Discovering regime-dependent governing equations via symbolic decision trees} \label{sec: model}

\begin{figure}[h]
    \centering
\begin{tikzpicture}[level/.style={sibling distance=30mm/#1}]
\node [circle,draw, very thick] at (0,0) (root){\Large{$1$}}
  child {node [circle,draw, very thick] (pow1) {\Large{$2$}}  }
  child {node [circle, draw, very thick] (mul) {\Large{$3$}}};
\node at (1.8,-0.5) {\Large{$g(x) \geq b_{1}$}};
\node at (-1.8,-0.5) {\Large{$g(x) < b_{1}$}};
\node at (-1.6,-2.25) {{$y_{2} = \sum_{k} c_{2,k} \phi_{k}(x)$}};
\node at (1.6,-2.25) {{$y_{3} = \sum_{k} c_{3,k} \phi_{k}(x)$}};
\end{tikzpicture}
    \caption{Decision trees with nonlinear expressions in the leaves and splitting logic.}
    \label{fig:decision tree}
\end{figure}
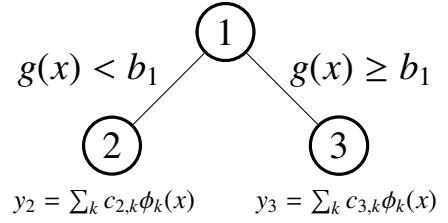

In this section, we present the mathematical model for learning piecewise continuous models. We assume that the model has $N_{x}$ input variables $x \in \mathbb{R}^{N_{x}}$ and one output $y \in \mathbb{R}$. The behavior of the system is governed by the following regime-dependent equation
\begin{equation}
    y(x)= \begin{cases}
f_{1}(x), & \text{if} \ \ x \in \mathcal{R}_{1} \\
\vdots\\
f_{n}(x), & \text{if} \ \ x \in \mathcal{R}_{n},
\end{cases}
\end{equation}
where $\mathcal{R}_{i} = \{x: g_{i}(x)\leq b_{i}\}$ denotes a regime described by a set of inequalities $g_{i}(x) \leq b_{i}$. We assume that the regimes are non-overlapping, i.e., a point $x$ belongs to at most one regime. We will parameterize the model using a decision tree. An example of a model with two regimes is presented in Fig~\ref{fig:decision tree}.

We assume that we are given dataset of the form $\mathcal{D} = \{x_{i},y_{i}\}_{i=1}^{N_{\rm{d}}}$, with $N_{\rm{d}}$ being the number of data points, and define as $\mathcal{I}$ the set of data points. We also define $D$ as the depth of the tree and $N = 2^{D+1}-1$ as the number of nodes in the tree. Each node $n$ has two children, nodes $2n$ and $2n+1$. We also define $\mathcal{N} = \{1,\dots, N\}$ as the set of nodes, $\mathcal{T} = \{2^{D},\dots, 2^{D+1}-1\}$ as the set of terminal nodes, and $\mathcal{N}_{int}=\mathcal{N}\setminus \mathcal{T}$ as the set of internal nodes. Finally, we define $\mathcal{K}_{b}$ and $\mathcal{K}_{f}$ as the set of basis functions used in the branching and local nonlinear expression in each leaf. In this decision tree representation of the model, each node $n$ is either a branching, a leaf, or an inactive node. 

\subsection{Tree structure constraints}
The first set of constraints ensures that the structure of the decision tree is correct. These constraints are adapted from \cite{bertsimas2017optimal,aghaei2025strong}. We define a binary variable $d_{n}$ which is equal to one if node $n$ is a branching node and zero otherwise. The logic condition determining whether a node $n$ can be a branching node is presented below
\begin{equation}\label{eq: tree 1}
    \begin{aligned}
        d_{2n} & \leq d_{n} && \ \forall n \in \mathcal{N}_{int}\\
        d_{2n+1} & \leq d_{n} && \ \forall n \in \mathcal{N}_{int}\\
        d_{1} & = 1 \\
        d_{n} & =0 && \ \forall n \in \mathcal{T}.
    \end{aligned}
\end{equation}
These constraints enforce that node $n$ can be a branching node only if the parent is also a branching node; otherwise, it must be inactive or a terminal node. For example, if node $n$ is a leaf node, i.e., $d_{n}=0$, then the children nodes can not be branching nodes or leaf nodes, i.e., $d_{2n}=0$ and $d_{2n+1}=0$. The last two constraints ensure that the root node is a branching node and that leaf nodes are either terminal or inactive. 

\subsection{Assigning data points to nodes}
We define a binary variable $z_{in}$ which is equal to one if data point $i$ is assigned to node $n$ and zero otherwise. The assignment of data points to nodes is constrained by the type of the node, i.e., terminal vs. branching, and the branching logic. First, we enforce that if a node is a branching node, i.e., $d_{n}=1$, then data can not be assigned to that node, i.e., $z_{in}=0$, as follows
\begin{equation}\label{eq: tree 2}
    z_{in} \leq 1-d_{n} \ \ \forall i \in \mathcal{I}, n \in \mathcal{N}.
\end{equation}
For each data point, the decision tree model must generate a prediction, i.e., each data point $i$ must be assigned to one node. This is enforced via the following constraints
\begin{equation}\label{eq: tree 3}
    \sum_{n \in \mathcal{N}} z_{in} = 1 \ \forall  i \in \mathcal{I}. 
\end{equation}
The assignment of a data point $i$ to a node $n$ can be allowed only if all the ancestors of node $n$ are branching nodes. This requirement is enforced as follows
\begin{equation}\label{eq: tree 4}
    z_{in} \leq d_{n'} \ \forall  i \in \mathcal{I}, n' \in A(n),
\end{equation}
and guarantees that data are not assigned to inactive or branching nodes. $A(n)$ denotes the ancestors of node $n$, i.e., all the nodes starting from the root until the parent of node $n$.

\subsection{Splitting logic}
We assume that the splitting condition at node $n$ can be written as $g_{n}(x) < b_{n}$ and $g_{n}(x) \geq b_{n}$, with $b_{n}$ being the threshold for node $n$. We model the expression $g_{n}(x)$ using basis functions as follows
\begin{equation*}
    g_{n}(x) = \sum_{k \in \mathcal{K}_{b}} a_{nk} \phi_{k}(x)\\
\end{equation*}
where $a_{nk} \in [a^{\rm{lb}}, a^{\rm{ub}}]$ is the coefficient for the $k^{th}$ basis function used in the splitting equation at node $n$. We define a binary variable $\omega_{nk}$ which is equal to one if feature $k$ is used in the splitting condition at node $n$ and zero otherwise. These variables constrain the value of the coefficients in the branching expression for each node as follows
\begin{equation}
    \begin{aligned}
        a_{nk} & \leq a^{\rm{ub}} \omega_{nk} && \forall n \in \mathcal{N}_{int}, k \in \mathcal{K}_{b}\\
        a_{nk} & \geq a^{\rm{lb}} \omega_{nk} && \forall n \in \mathcal{N}_{int}, k \in \mathcal{K}_{b}.
    \end{aligned}
\end{equation}
If node $n$ is not a branching node, i.e., $d_{n}=0$, then all the constants related to the branching expression must be zero. This is enforced via the following constraint
\begin{equation}
    \omega_{nk} \leq d_{n} \ \ \forall  \ n \in \mathcal{N}_{int}, k \in \mathcal{K}_{b}.
\end{equation}
The last set of constraints guarantees that the assignment of data point $i$ to node $n$ satisfies the branching logic. Specifically, if data point $i$ is assigned to node $n$, i.e., $z_{in}=1$, then it must satisfy all the logic conditions that hold at the ancestors of node $n$. We define as $R(n)$ and $L(n)$ the nodes that are in the path from the root node to node $n$, which are branched right and left, respectively. The routing constraints are
\begin{equation}\label{eq: routing}
    \begin{aligned}
        \sum_{k\in \mathcal{K}_{b}} a_{mk} \phi_{k}(x_{i}) & \leq b_{m} - \epsilon + M(1-z_{in}) && \forall \ i \in \mathcal{I}, n \in \mathcal{N}, m \in L(n)\\
        \sum_{k\in \mathcal{K}_{b}} a_{mk} \phi_{k}(x_{i}) & \geq b_{m} - M(1-z_{in})  && \forall \ i \in \mathcal{I}, n \in \mathcal{N}, m \in R(n).
    \end{aligned}
\end{equation}

\subsection{Computing the output of the model for regression tasks}
This set of constraints concerns the computation of the model's output. We define $c_{kn} \in [c^{\rm{lb}}, c^{\rm{ub}}]$ as the constant for basis function $k$ at node $n$ used to predict the output of the model. We also define $\hat{y}_{in} \in [y^{\rm{lb}}, y^{\rm{ub}}]$ as the prediction of the expression at node $n$ for data point $i$. The prediction at each node and data point is computed by the following constraints
\begin{equation}
    \hat{y}_{in} = \sum_{k \in \mathcal{K}_{f}} c_{kn} \phi_{k}(x_{i}) \ \forall i \in \mathcal{I}, n \in \mathcal{N},
\end{equation}
where $\phi_{k}(x_{i})$ is the $k^{th}$ basis function for data point $i$. We define a binary variable $w_{nk}$ which is equal to one if the basis function $k$ is used at the local expression at node $n$ and zero otherwise. The constants in the basis functions should be equal to zero if the node is a branching node. This is enforced via the following constraints
\begin{equation}\label{eq: select constants nonlinear prediction}
\begin{aligned}
    c_{kn} \leq c^{\rm{ub}} w_{nk} && \forall k \in \mathcal{K}_{f}, n \in \mathcal{N}\\
    c_{kn} \geq c^{\rm{lb}} w_{nk} && \forall k \in \mathcal{K}_{f}, n \in \mathcal{N}\\
    w_{nk} \leq 1-d_{n} && \forall k \in \mathcal{K}_{f}, n \in \mathcal{N}
\end{aligned}
\end{equation}
We also define $y_{i}^{\rm{pred}} \in [y^{\rm{lb}}, y^{\rm{ub}}]$ as the prediction of the decision tree for data point $i$ which is equal to
\begin{equation}
    y_{i}^{\rm{pred}} = \sum_{n \in \mathcal{N}} z_{in} \hat{y}_{in}.
\end{equation}
This constraint has bilinear terms between a continuous $\hat{y}_{in}$ and a binary $z_{in}$ variable, which can be linearized exactly. We define $\delta_{in} \in [y^{\mathrm{lb}},y^{\mathrm{ub}}]$ with $\delta_{in} = \hat{y}_{in}z_{in}$. Each bilinear term is linearized as follows
\begin{equation}\label{eq: predict 4}
    \begin{aligned}
        & \delta_{in} \leq y^{\mathrm{ub}} z_{in}\\
        & \delta_{in} \geq y^{\mathrm{lb}} z_{in}\\
        & \delta_{in} \leq \hat{y}_{in} + M (1 - z_{in})\\
        & \delta_{in} \geq \hat{y}_{in} - M (1 - z_{in}),
    \end{aligned}
\end{equation}
where $M$ is a large constant. Under this linearization, we have
\begin{equation}\label{eq: predict 6}
    y^{\rm{pred}}_{i} = \sum_{n \in \mathcal{N}} \delta _{in} \ \forall i \in \mathcal{I}
\end{equation}

\subsection{Constraining the complexity of the branching logic and nonlinear governing equations}
The last set of constraints is used to control the complexity of the branching and local nonlinear expressions. We define $N_{\rm{B}}$ as the maximum number of features that can be used in a branching node and $N_{\rm{F}}$ as the maximum number of features that can be used in each nonlinear expression at the leaves. The maximum number of features that can be used for branching and predicting the output at each node is bounded as follows 
\begin{equation}\label{eq: complexity branching}
    \begin{aligned}
        & \sum_{ k \in \mathcal{K}_{b}} \omega_{nk} \leq N_{\rm{B}} \ \ \forall n \in \mathcal{N}_{int}.
    \end{aligned}
\end{equation}
\begin{equation}\label{eq: complexity prediction}
    \begin{aligned}
        & \sum_{ k \in \mathcal{K}_{f}} w_{nk} \leq N_{\rm{F}} \ \ \forall n \in \mathcal{N}.
    \end{aligned}
\end{equation}

\subsection{Learning task for regression}

The objective of the learning problem has three components: prediction error $\mathcal{L}_{loss}$, complexity of the tree $\mathcal{L}_{c}$, and magnitude of the constants $\mathcal{L}_{c}$. Each term is computed as follows
\begin{equation*}
    \begin{aligned}
        \mathcal{L}_{acc} & = \frac{1}{N_{\rm{d}}}\sum_{i \in \mathcal{I}} |y_{i} - y^{pred}_{i}|\\
        \mathcal{L}_{c} & = \sum_{n \in \mathcal{N}} d_{n}\\
        \mathcal{L}_{m} & = \sum_{k \in \mathcal{K}} \sum_{n \in \mathcal{N}} |c_{kn}|. 
    \end{aligned}
\end{equation*}
We reformulate the absolute terms by defining the following nonnegative variables $\epsilon^{+}_{i}, \epsilon_{i}^{-}$, $\hat{c}_{kn}^{+}$, $\hat{c}_{kn}^{-}$
and writing the absolute values as follows
\begin{equation}\label{eq: predict 6}
    \begin{aligned}
         \epsilon_{i}^{+}-\epsilon_{i}^{-} & = y_{i}-y_{i}^{pred} \ \ && \forall i \in \mathcal{I}\\
         \hat{c}_{kn}^{+}-\hat{c}_{kn}^{-} & = c_{kn} \ \ && \forall k \in \mathcal{K}_{f}, n \in \mathcal{N}.
    \end{aligned}
\end{equation}
The reformulated $\mathcal{L}_{acc}$ and $\mathcal{L}_{m}$ terms are
\begin{equation*}
    \begin{aligned}
        \mathcal{L}_{acc} & = \frac{1}{N_{\rm{d}}}\sum_{i \in \mathcal{I}} \big(\epsilon_{i}^{+} + \epsilon_{i}^{-}\big)\\
        \mathcal{L}_{m} & = \sum_{k \in \mathcal{K}} \sum_{n \in \mathcal{N}} \big(\hat{c}_{kn}^{+} + \hat{c}_{kn}^{-}\big),
    \end{aligned}
\end{equation*}
and the learning problem is
\begin{equation}
    \begin{aligned}
        \min \ \ & \mathcal{L}_{acc} + \lambda_{c} \mathcal{L}_{c} + \lambda_{m} \mathcal{L}_{m}\\
        \text{s.t.} \ \ & \text{Eq.}~\ref{eq: tree 1}-\ref{eq: predict 6},
    \end{aligned}
\end{equation}
where $\lambda_{c}$, $\lambda_{m}$ are weight factors for the different terms in the objective. Overall, the learning problem is a mixed integer linear optimization problem that can be solved with existing optimization solvers.

\begin{remark}
    \normalfont The formulation presented above can be readily extended for classification tasks. For the case of binary classification, following the formulation of the optimal trees \cite{bertsimas2017optimal,aghaei2025strong}, the prediction at each node will be a binary variable and the prediction error term $\mathcal{L}_{acc}$ will be the number of misclassified data points. 
\end{remark}

\section{Case study 1: Illustrative example} \label{sec: simple example}
First, we consider the illustrative example presented below
\begin{equation}\label{eq: case 1 true model}
    y(x_{1},x_{2}) = \begin{cases}
    \begin{aligned}
        x_{1}^{2}+x_{2}^{2} \ \ & \text{if} \ \ x_{1}^{2}+x_{2}^{2}\leq 2.5\\
        x_{1}^{2}+x_{2} \ \ & \text{otherwise}.
        \end{aligned}
    \end{cases}
\end{equation}
We generate training data for $x_{1} \in [-2,2]$ and $x_{2} \in [-2,2]$ by uniformly sampling $N=100$ points. The basis functions for the branching expressions are 
\[ \{x_{1}, x_{2}, x_{1}^{2}, x_{2}^{2}, x_{1}x_{2}\}\]
and for the leaf nodes 
\[\{1, x_{1}, x_{2},x_{1}^{2},x_{2}^{2}, x_{1}x_{2}\}.\] 
Regarding the model parameters, we set $M=100$, $\lambda_{c}=0$, $\lambda_{m}=0$, $a^{\rm{lb}}=-10^{2}$,  $a^{\rm{ub}}=10^{2}$, $b^{\rm{lb}}=-10^{2}$, $b^{\rm{ub}}=10^{2}$, $c^{\rm{lb}}=-10^{3}$, $c^{\rm{ub}}=10^{3}$, $y^{\rm{lb}}=-10^{3}$, $y^{\rm{ub}}=10^{3}$, and $N_{b}=2$. For this case, we do not include the binary variables $w_{kn}$ and the associated constraints in Eq.~\ref{eq: select constants nonlinear prediction} and \ref{eq: complexity prediction}. The learning task has 1268 variables (307 binary) and 2572 constraints, and is solved in 5 seconds using Gurobi 11.0.0 \cite{gurobi}. 

The discovered regime-dependent governing equation is
\begin{equation} \label{eq: case 1 learned model}
    \begin{cases}
    \begin{aligned}
        x_{1}^{2}+x_{2}^{2} \ \ & \text{if} \ \ 1.01 x_{1}^{2}+x_{2}^{2}\leq 2.54\\
        x_{1}^{2}+x_{2} \ \ & \text{otherwise}.
        \end{aligned}
    \end{cases}
\end{equation}
Comparing the learned and exact equation (Eqs.~\ref{eq: case 1 true model} and \ref{eq: case 1 learned model}), we observe that the main difference is in the coefficients of the splitting equations; the expressions in each regime are identified correctly since there is no noise in the training data. The coefficients in the regime boundary equation differ due to the limited data.

\begin{figure}[h]
    \centering
    \includegraphics[width=\linewidth]{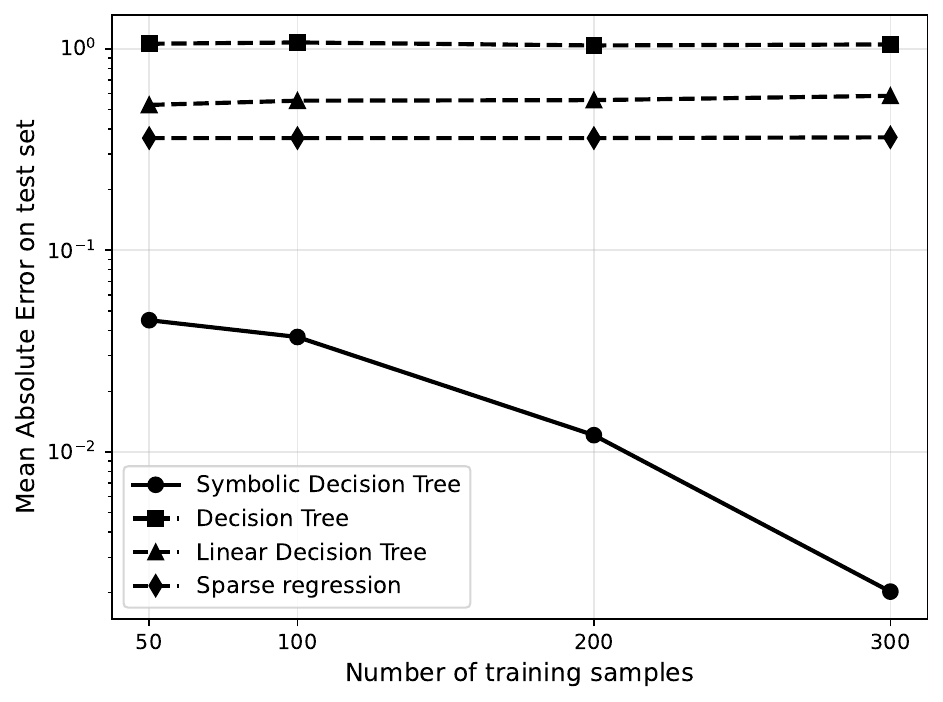}
    \caption{Mean absolute error on the testing set as a function of the number of data in the training set for the different models.}
    \label{fig:error simple case}
\end{figure}

\begin{figure*}[t]
    \centering
    \includegraphics[width=\linewidth]{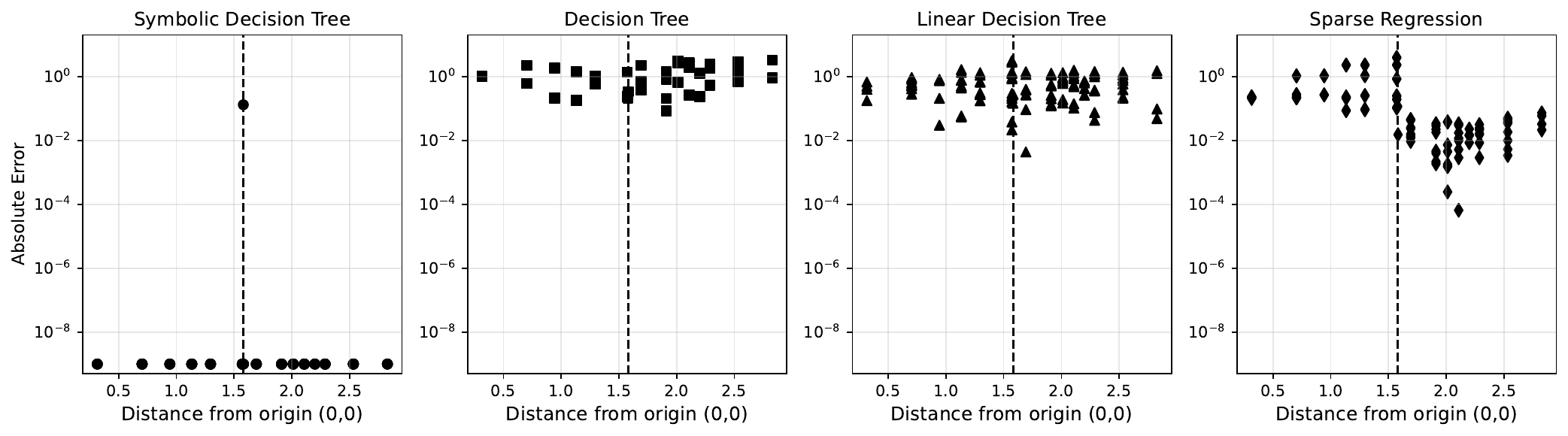}
    \caption{Mean absolute error on the testing set as a function of the distance from the origin for the different models.}
    \label{fig:error vs distance simple case}
\end{figure*}

We also compare the proposed method with three baselines: sparse regression, decision trees with constant predictions at the leaves, and decision trees with linear expressions at the leaves. For the sparse regression, we use the same basis functions as for the symbolic decision trees and for the decision-tree models we fix the depth of the tree to one. The sparse regression problem is formulated as a linear optimization problem that minimizes the $\mathcal{L}_{\rm{acc}}$ loss term and is solved with Gurobi. The decision tree model with constant predictions is implemented in scikit-learn \cite{scikit-learn} and the one with the linear prediction in lineartree Python package.

The predictive accuracy of each method on the testing set for different number of training data is presented in Fig.~\ref{fig:error simple case}. From the figure, we observe that the mean absolute error of the proposed approach is always lower than that of the baselines, and decreases as the number of training data points increases. Moreover, we observe that the standard decision trees with constant predictions at the leaves have the highest error, followed by decision trees with linear leaves and sparse regression. For all the baselines, the MAE remains constant as the number of training data points increases, since these models can not identify the change in the governing equations.

Finally, we analyze the error of the learned symbolic decision tree model as a function of the distance from the origin. We uniformly discretize the domain of each variable into 10 points and compute the absolute error for the proposed approach and the baselines. The absolute error as a function of the distance from the origin is presented in Fig.~\ref{fig:error vs distance simple case}. We observe that the error of the proposed model is highest near the boundary and zero elsewhere, since the governing equation for each regime is correctly identified. This is due to limited training data, which creates degeneracy in the learning task. 

Compared to the baselines, we observe that the model learned with the proposed approach has lower prediction error. Specifically, the sparse regression approach has higher accuracy in the regime outside the cycle and yields larger errors for points inside the circle. Finally, both decision tree models, the one with constant predictions and the one with linear predictions, have similar error across the entire domain. Overall, these results show that the proposed approach can identify the partition of the input space into regions and the corresponding model within each region. 

\begin{figure}[h]
    \centering
    \includegraphics[width=\linewidth]{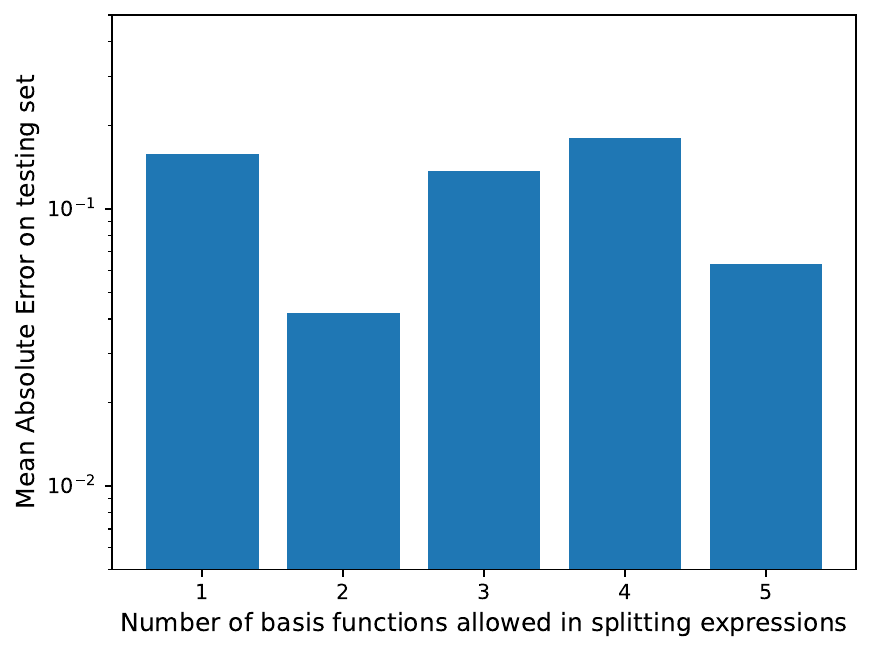}
    \caption{Mean absolute error on the testing set as a function of the maximum number of features that can be used in the splitting conditions.}
    \label{fig:MAE vs N split case 1}
\end{figure}

Finally, we note that in this case, the maximum number of features that can be used in the branching equations was set equal to two. If the maximum number of features is not specified, the solution of the model might return a splitting condition that does not correspond to the true splitting conditions. This is due to the inherent degeneracy in such problems and overfitting. Due to limited training data, two partitions can yield the same objective function value if the local models are correctly identified. However, the optimal value can be found by computing the mean absolute error (MAE) on the testing set. For this case study, the MAE on the testing set for varying maximum number of basis functions is presented in Fig.~\ref{fig:MAE vs N split case 1}. The results show that the lowest MAE is achieved if at most two features are used at the branching nodes ($N_{\rm{B}}=2$).

\section{Case study 2: Learning hybrid dynamical models}\label{sec: learn hybrid models}

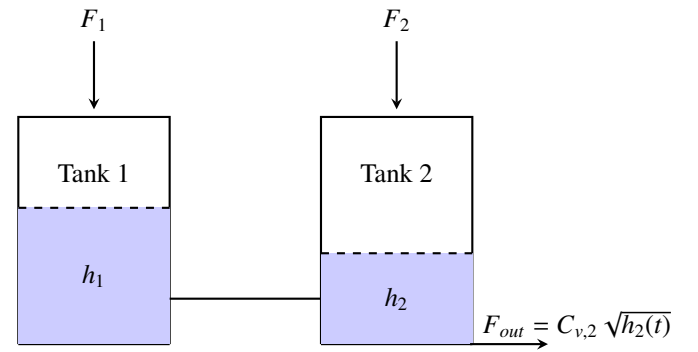
\begin{figure}[h]
\centering
\begin{tikzpicture}[>=stealth, thick]

    % --- Tank 1 ---
    \draw (0,0) -- (0,3) -- (2,3) -- (2,0);          % walls (no bottom drawn)
    \draw (0,0) -- (2,0);                              % bottom
    \fill[blue!20] (0,0) rectangle (2,1.8);            % liquid level h1
    \draw[dashed] (0,1.8) -- (2,1.8);                  % h1 level line
    \node at (1, 0.9) {$h_1$};
    \node[above] at (1, 2) {Tank 1};

    % --- Tank 2 ---
    \draw (4,0) -- (4,3) -- (6,3) -- (6,0);
    \draw (4,0) -- (6,0);
    \fill[blue!20] (4,0) rectangle (6,1.2);            % liquid level h2
    \draw[dashed] (4,1.2) -- (6,1.2);                  % h2 level line
    \node at (5, 0.6) {$h_2$};
    \node[above] at (5, 2) {Tank 2};
    % --- Connecting pipe between tanks ---
    \draw (2,0.6) -- (4,0.6);                          % pipe
    % \node[above] at (3, 0.6) {$C_v$};
    % flow direction arrow (h1 > h2 case)
    % \draw[->] (2.6,0.6) -- (3.4,0.6);
    % \node[below, font=\small] at (3, 0.6) {$h_1>h_2$};

    % --- Inlet F1 ---
    \draw[->] (1, 4) -- (1, 3.1);
    \node[above] at (1, 4) {$F_1$};

    % --- Inlet F2 ---
    \draw[->] (5, 4) -- (5, 3.1);
    \node[above] at (5, 4) {$F_2$};

    % --- Outlet from Tank 2 ---
    \draw[->] (5.97, 0) -- (7, 0);
    % \draw[->] (5.5, -0.6) -- (5.5, -1.1);
    \node[right] at (6, 0.25) {$F_{out} = C_{v,2} \sqrt{h_{2}(t)}$};
    % \node[right] at (5.5, -0.8) {$F_{out},\ C_{v,2}$};

\end{tikzpicture}
\caption{Schematic of the interacting two-tank system.}
\label{fig:two_tank_schematic}
\end{figure}

In this section, we consider the identification of hybrid dynamical systems, using an interacting two-tank system as a case study, as shown in Fig.~\ref{fig:two_tank_schematic}. 

The dynamic behavior of the liquid level in the first tank is described by the following ordinary differential equation
\begin{equation} \label{eq: tank 1 model}
\begin{aligned}
        \frac{dh_{1}(t)}{dt} & = \begin{dcases}
        \frac{F_{1}(t)- C_{v} \sqrt{|h_{1}(t)-h_{2}(t)|}}{A_{1}} \ \ \text{if} \ \ h_{1}(t)-h_{2}(t)>0 \\
        \frac{F_{1}(t)+ C_{v} \sqrt{|h_{1}(t)-h_{2}(t)|}}{A_{1}} \ \ \text{if} \ \ h_{1}(t)-h_{2}(t) \leq 0
    \end{dcases}
\end{aligned}
\end{equation}
\begin{equation}
\begin{aligned}
    \frac{dh_{2}(t)}{dt} & = \begin{dcases}
        \text{if} \ \ h_{1}(t) >h_{2}(t)  \\
         \frac{F_{2}(t)+ C_{v} \sqrt{|h_{1}(t)-h_{2}(t)|} - C_{v,2} \sqrt{h_{2}(t)}}{A_{1}}  \\
         \text{if} \ \ h_{1}(t)-h_{2}(t) \leq 0\\
        \frac{F_{2}(t)- C_{v} \sqrt{|h_{1}(t)-h_{2}(t)|} - C_{v,2} \sqrt{h_{2}(t)}}{A_{1}},
    \end{dcases}
\end{aligned}
\end{equation}
where $h_{1}(t)$ is the liquid level, $h_{2}(t)$ the liquid level in the second tank, $F_{1}(t)$ is the inlet flow rate for the first tank, $F_{2}(t)$ the inlet flow rate for the second tank, 
$A_{1}=1$ is the cross-sectional area of the first tank, and $C_{v}=C_{v,2}=0.5$. 
\begin{figure}[h]
    \centering    \includegraphics[width=\linewidth]{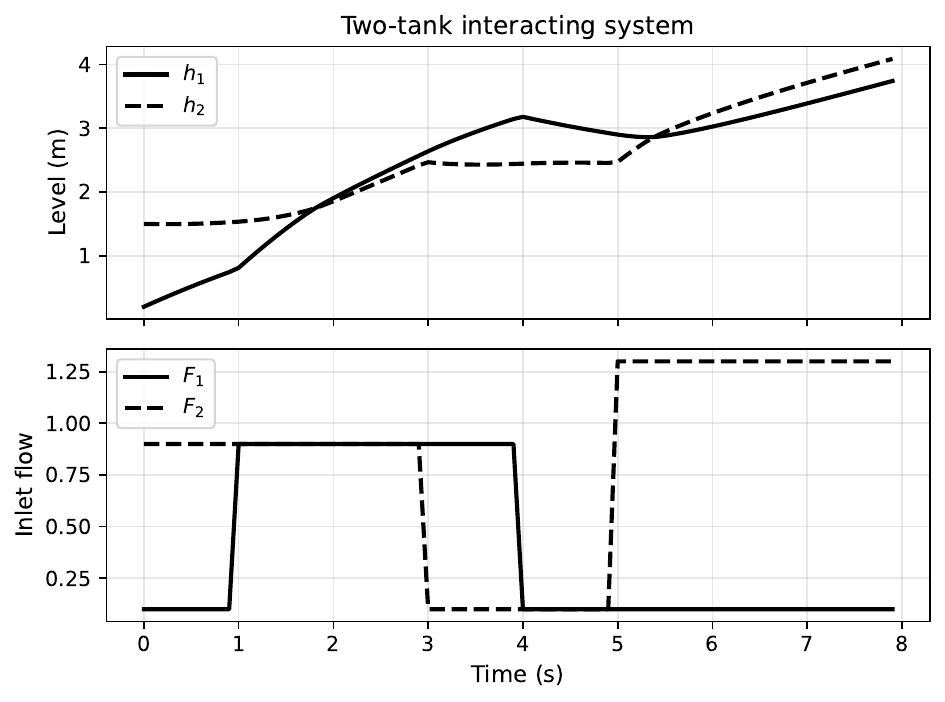}
    \caption{Inlet flow rates and liquid level evolution}
    \label{fig:case 2 training data}
\end{figure}

\begin{figure*}[h]
    \centering
    \includegraphics[width=\linewidth]{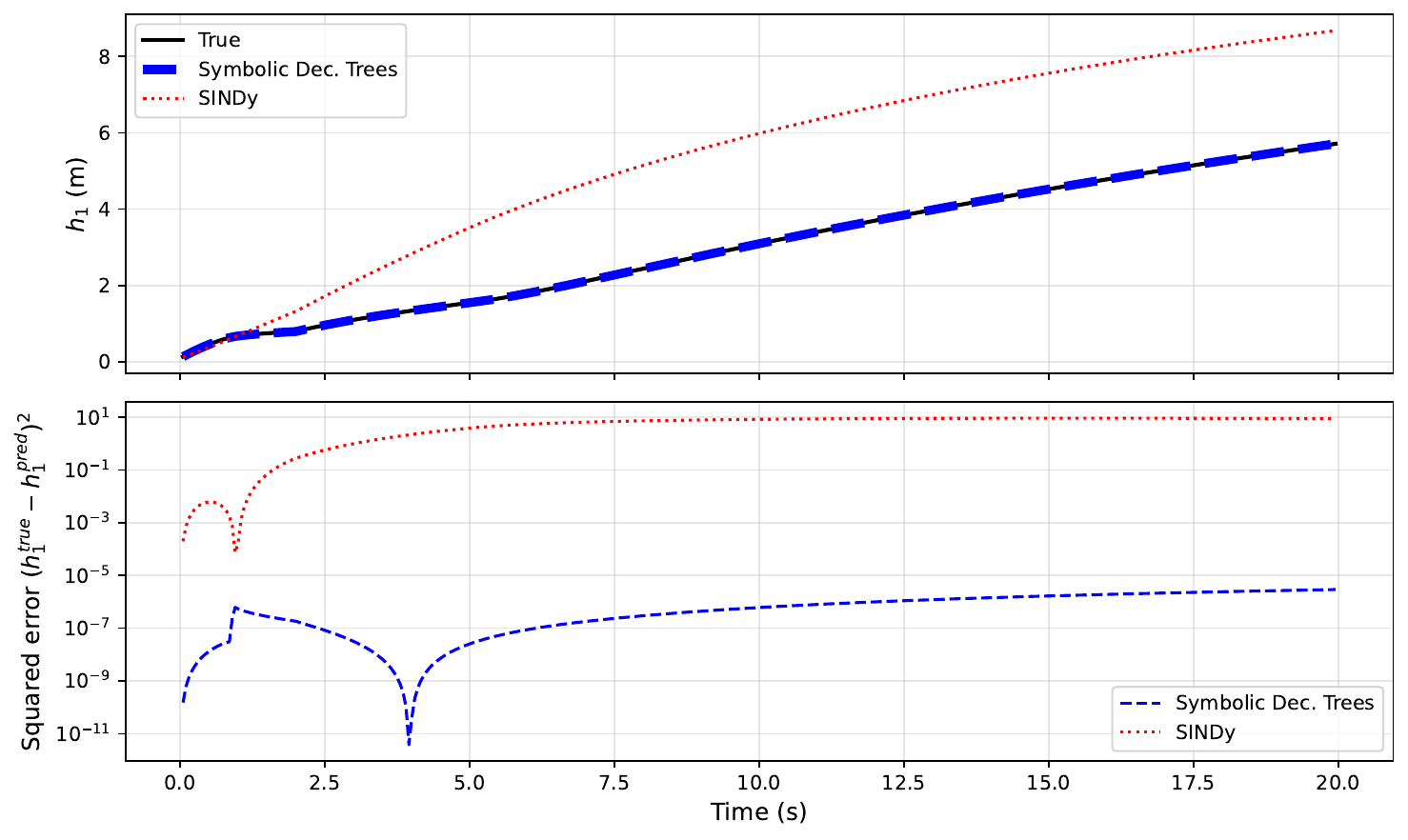}
    \caption{Evolution of the liquid level in the first tank using the model discovered with the proposed approach and with sparse regression.}
    \label{fig:case 2 test trajectory}
\end{figure*}
We generate one training trajectory from the initial condition $[h_{1}(0) \ \ h_{2}(0)]=[0.2 \ \ 1.5]$ and change the inlet flow rates as presented in Fig.~\ref{fig:case 2 training data}. We integrate the system for $t \in [0,8]$ using the $4^{\rm{th}}$ order Runge-Kutta method with a discretization step equal to 0.1. The basis functions for the branching conditions are
\begin{equation*}
    [ h_{1} - h_{2} , h_{1}, h_{2}, F_{1}]
\end{equation*}
and for the local nonlinear expressions
\begin{equation*}
    [1 , \sqrt{|h_{1}-h_{2}|}, \sqrt{h_{2}}, F_{1}].
\end{equation*}
We use these basis functions since we only want to learn the differential equation for the first tank. We also set $N_{\rm{B}}=1$ and $N_{\rm{F}}=2$ and the depth of the tree is 1. The learning task has 2066 constraints and 1019 variables, including 258 binary. 

The learned model for the first tank is
\begin{equation}
\begin{aligned}
        \begin{cases}
         -0.5\sqrt{|h_{1}(t)-h_{2}(t)|} + F_{1}(t) \ \ \text{if} \ \ h_{1}(t)-h_{2}(t)\geq -0.002  \\
        0.5\sqrt{|h_{1}(t)-h_{2}(t)|} + F_{1}(t)\ \ \text{otherwise}.
    \end{cases}
\end{aligned}
\end{equation}
Comparing the learned and true model (Eq.~\ref{eq: tank 1 model}), we observe that the governing equations for each regime are identified correctly. The main difference lies in the splitting conditions, where the right-hand side equals $-0.002$ rather than zero. 

We compare the learned model with sparse regression a widely used approach for discovering dynamical models. We use the same basis functions as the ones used for the leaf nodes of the proposed symbolic decision tree model. The model discovered with sparse regression is
\begin{equation}
    \frac{dh_{1}(t)}{dt}= 0.716  -0.021 \sqrt{|h_{1}-h_{2}|} -0.312 \sqrt{h_{2}} + 0.597 F_{1}.
\end{equation}
Although this model can not identify the regime-dependent equations by construction, we observe that the discovered model does not correspond the governing equations of either domain. 

We also compare the predictive accuracy of the discovered models for a test trajectory, and the results are presented in Fig.~\ref{fig:case 2 test trajectory}. We simulate the system for $t \in [0,20]$ using $[h_{1}(0) \ \ h_{2}(0)]=[0.1 \ \ 1.2]$ as initial condition. From the results, we observe that the model discovered with the proposed approach tracks the evolution of height more accurately than the model discovered using sparse regression. Specifically, the root mean squared error with the proposed approach is $9.6 \times 10^{-4}$, whereas with the sparse regression model it is 2.53. 

We note that we do not compare the proposed approach with existing two-step approaches that first partition the regime and then fit the model, since in such cases, the splitting logic is not interpretable, and the identified partition is not necessarily optimal for model accuracy.

\section{Case study 3: Learning molecular-weight scaling laws for polymer zero-shear viscosity} \label{sec: predict polymer property}

Finally, we consider the discovery of constitutive relationships to predict the zero-shear viscosity of polymers as a function of molecular weight. To describe the crossover from unentangled to entangled dynamics, we use a piecewise scaling model that combines Rouse-like behavior at low molecular weight and entangled-polymer scaling at high molecular weight:
\begin{equation}
\begin{aligned}
        \frac{\eta_{0}}{\eta_{c}} = \begin{cases}
        \frac{M}{M_{c}} \ \ & \text{if} \ \ M< M_{c}\\
        (\frac{M}{M_{c}})^{3.4} \ \ & \text{if} \ \ M \geq M_{c},
    \end{cases}
\end{aligned}
\end{equation}
where $M$ is the molecular weight, $M_{c}$ is the critical molecular weight for the onset of entanglement effects, $\eta_{0}$ is the zero-shear viscosity, and $\eta_{c}$ is the zero-shear viscosity at $M = M_{c}$. 

\begin{figure}[t]
    \centering
    \includegraphics[width=\linewidth]{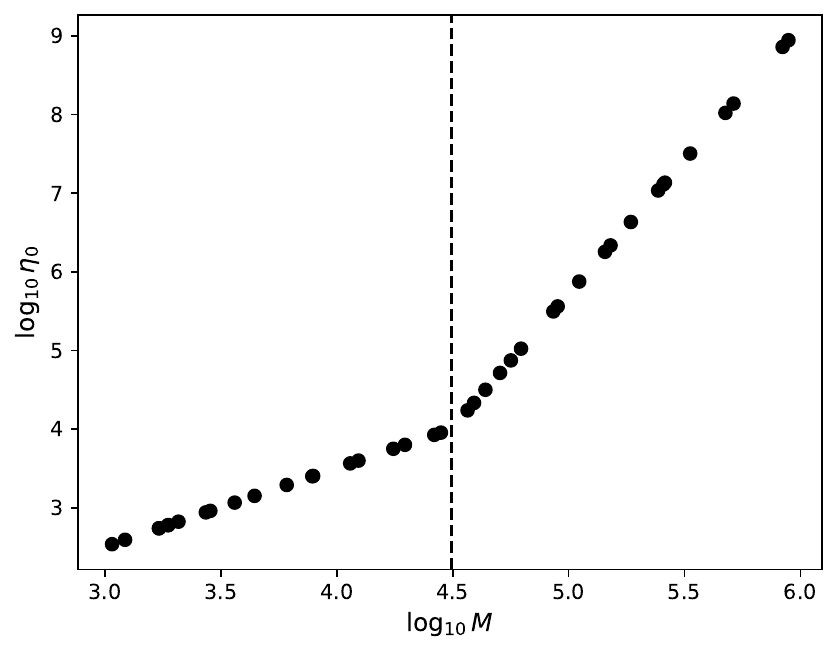}
    \caption{Polymer viscosity as a function of molecular weight for $M_{c}=31200$ and $\eta_{c}=10^{4}$.}
    \label{fig:viscosity plot}
\end{figure}

We generate training data by sampling 40 molecular weights over the range $(10^{3}, 10^{6})$, setting $M_{c} = 31200$ and $\eta_{c}=10^{4}$ (Fig.~\ref{fig:viscosity plot}). Molecular weights are sampled uniformly in logarithmic scale to provide comparable resolution across the unentangled and entangled regimes. Because polymer viscosity data are typically analyzed on a logarithmic scale, the training data are transformed using $\log_{10}\eta_0$. The resulting piecewise model is
\begin{equation} \label{eq: viscosity model}
\begin{aligned}
        \log_{10} \eta_{0} = \begin{cases}
        \log_{10} M -0.49 & \text{if} \  \log_{10} M < 4.49\\
        3.4 \log_{10} M -11.28 & \text{if} \ \ \log_{10} M \geq 4.49.
    \end{cases}
\end{aligned}
\end{equation}

For the learning task, we use $[\log_{10} M, M]$ as candidate basis functions for the splitting conditions, $[1, \log_{10} M, M]$ as candidate basis functions for the local nonlinear expressions, and $1$ as the maximum tree depth. The resulting optimization problem contains 1036 constraints and 514 variables, including 124 binary variables. The optimal solution corresponds to the following piecewise equation
\begin{equation}\label{eq: learned viscosity model}
\begin{aligned}
        \log_{10} \eta_{0} = \begin{cases}
        \log_{10} M -0.49 & \text{if} \  \log_{10} M < 4.24\\
        3.4 \log_{10} M -11.28 & \text{if} \ \ \log_{10} M \geq 4.24.
    \end{cases}
\end{aligned}
\end{equation}
Comparing this learned model with the true model in Eq.~\ref{eq: viscosity model}, we find that the algorithm correctly identifies the functional form of the viscosity expression in each regime. Specifically, the learned model recovers the expected slopes of 1 and 3.4 in the low and high molecular weight regimes, respectively, corresponding to Rouse-like and entangled-polymer scaling. The primary difference is the location of the splitting condition, which occurs at $\log_{10}M = 4.24$ in the learned model instead of $\log_{10}M = 4.49$ in the true model. In addition, whereas the true model is continuous across the crossover, the discovered model is discontinuous because the regime-dependent governing equation is represented using a decision tree. Increasing the number of training points from 40 to 100 yields the same regime-specific functional forms and improves the learned splitting condition to $\log_{10}M = 4.45$. Thus, the model captures the underlying scaling laws, while the inferred crossover location becomes more accurate with increased data density. 
\begin{figure}[h]
    \centering
    \includegraphics[width=\linewidth]{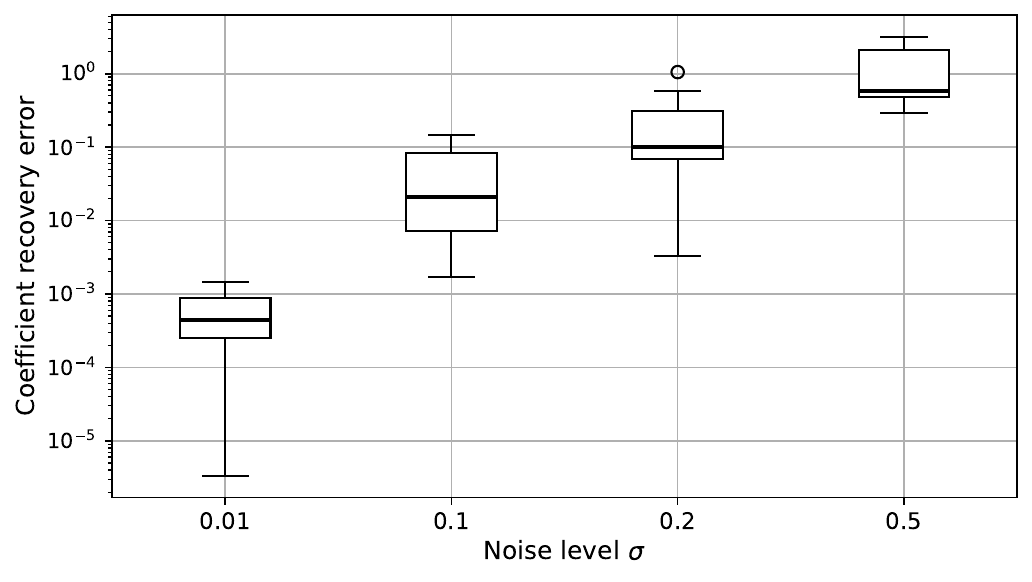}
    \caption{$L_{2}$ error for the identified coefficients as a function of the noise level.}
    \label{fig:error robustness}
\end{figure}

\begin{figure}[h]
    \centering
    \includegraphics[width=\linewidth]{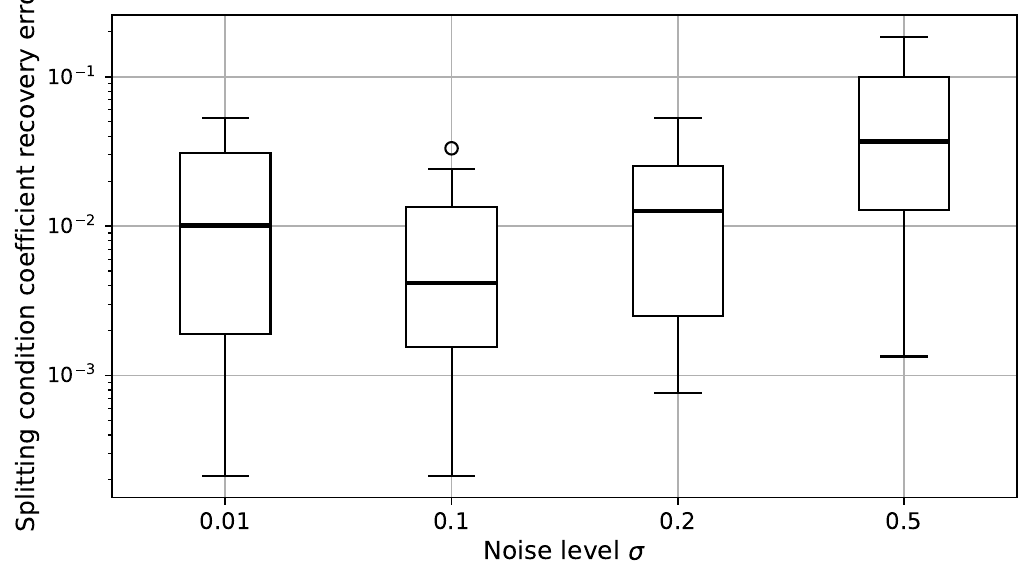}
    \caption{$L_{2}$ error for the identified coefficients in the splitting conditions as a function of the noise level.}
    \label{fig:error robustness splitting condition}
\end{figure}

Finally, we evaluate the robustness of the proposed approach to noise in the training data. Gaussian noise with zero mean and standard deviation $\sigma$ is added to the viscosity data, and the coefficient recovery error for the local expressions is quantified using the $L_{2}$ norm over 10 random seeds. The coefficient error increases with increasing noise level (Fig. ~\ref{fig:error robustness}); however, the learned equations retain the correct functional form, demonstrating that the approach robustly identifies the underlying regime-dependent viscosity model. By contrast, the coefficients of the splitting conditions are comparatively insensitive to noise (Fig.~\ref{fig:error robustness splitting condition}).

\section{Conclusions}
In this paper, we propose a data-driven framework for discovering regime-dependent governing equations. The major challenge in such learning tasks is the need to simultaneously identify distinct regimes and the governing equations within each regime. To address this challenge, we represented regime-dependent governing equations as symbolic decision trees with nonlinear splitting conditions and nonlinear expressions at the leaves. We used basis functions to parameterize both the splitting conditions and the local governing expressions, thereby improving the tractability of the learning task while preserving interpretability. We demonstrated the proposed approach on hybrid dynamical models and on the discovery of polymer zero-shear viscosity scaling laws as a function of molecular weight. 

The results show that the proposed approach can lead to regime-dependent models with high predictive accuracy. For the case studies considered, the coefficient error for the basis functions in the splitting conditions is higher than that for the local nonlinear expressions. This behavior can be attributed to degeneracy in the learning task caused by limited training data. In tree-based models, several partitions of the input domain can yield the same predictive accuracy; that is, multiple models can achieve the same globally optimal solution in terms of the objective function value. However, global optimization algorithms typically stop once a model satisfies the desired optimality gap tolerance. Therefore, the identified splitting conditions can differ from the true regime boundaries, even when the learned local governing equations and overall predictions remain accurate.

The scalability of the proposed approach depends on the number of data points, the depth of the decision tree, and the set of basis functions used to parametrize the splitting conditions and local nonlinear expressions. Although this issue is common in optimization-based training of decision-tree models, several methods can be used to reduce the computational time. For example, one can exploit the underlying structure of the problem using decomposition-based optimization algorithms \cite{firat2020column, patel2024improved, aghaei2025strong, keegan2025acceleration} or improve scalability through tailored branching logic and/or formulations \cite{mazumder2022quant, hua2022scalable, heredia2025rs}. 

The ability of the proposed approach to discover the true splitting logic and local governing equations also depends on the choice of basis functions. This is a common limitation of basis-function-based learning approaches. Finally, because the returned equations are linear combinations of predefined basis functions, the framework is currently limited in its ability to discover fully general equation forms.

\section*{Acknowledgements}

IM and GES acknowledge financial support from the McKetta Department of Chemical Engineering at The University of Texas at Austin.

\section*{Data Availability}
Codes for computational experiments can be accessed via 
\href{https://github.com/PSE-Lab/Learning_regime_dependent_governing_equations_via_symbolic_decision_trees}{GitHub}.

\addcontentsline{toc}{section}{References}
\bibliographystyle{ieeetr}
\bibliography{refs}

\end{document}